# AESTHETIC PERSPECTIVES IN INFORMATION SYSTEMS RESEARCH: A HERMENEUTIC ANALYSIS

*Completed Research Paper*

Angelina Chen, University of Sydney, Sydney, Australia, **angelina.chen@sydney.edu.au**

Rick Sullivan, HEC Montréal, Montréal, Canada, **rick.sullivan@hec.ca**

Raffaele Ciriello, University of Sydney, Sydney, Australia, **raffaele.ciriello@sydney.edu.au**

## Abstract

*How might implicit aesthetic perspectives shape what Information Systems (IS) scholarship recognises as worthy of study (or not)? In this hermeneutic literature analysis, we surface foundational aesthetic assumptions underpinning IS research. We identify four perspectives (aesthetics as imitation, sensory experience, world-making, and political doing) that guide how IS scholars perceive and appreciate sociotechnical phenomena. These perspectives influence what becomes recognisable as legitimate research and what remains unseen. By making aesthetic assumptions explicit, we show how they form epistemic infrastructure that conditions horizons of inquiry. We apply this framework to algorithmic management and digitally mediated intimacy, revealing how alternative perspectives open new research questions whilst exposing dimensions that dominant framings overlook. This analysis foregrounds the importance of aesthetic philosophy to IS literature, offering a vocabulary for articulating how aesthetic perspectives shape theorising, method, and contribution.*

> *"In that moment of aesthetic contemplation, we are freed from the relentless striving of the will.*
> *We enter pure perception, losing ourselves in intuition, forgetting individuality, dissolving all relations of reason.*
> *The object becomes the Idea of its kind, and the knowing subject rises to the pure, will-less standpoint of contemplation;*
> *both now stand outside the stream of time and all relations.*
> *It no longer matters, then, whether one sees the sunset from a prison or palace."*
>
> *Arthur Schopenhauer, The World as Will and Representation, §38 (own meaning-preserving translation)*

## 1 Introduction

Mainstream Information Systems (IS) scholarship has traditionally framed digital technologies as instruments for enhancing organisational performance and managerial utility (Clarke & Davison, 2020). Whilst this instrumental perspective has contributed valuable insights, the emphasis on metrification and optimisation has, at times, limited the scope of inquiry by backgrounding the richness of lived experiences and the generative aspects of affect, intuition, and imagination. This is problematic because digital technologies are now de-facto sociopolitical infrastructures that shape attention, identity, affect, and sense-making (Kudina, 2021). They condition not only what actors do but how they perceive and value the world: algorithmic curation shapes worldviews, platforms mediate identity and emotion, remote-work infrastructures reconfigure the sense of physical presence and professional selfhood, and generative AI increasingly participates in intimacy and public discourse (Pariser, 2011; Prester et al., 2023; Riemer & Peter, 2021; Susskind, 2018). These developments raise questions of meaning and flourishing alongside efficiency and utility. As digital infrastructures become central to wellbeing and public life, it is time for the IS discipline to reconsider what it foregrounds and what such foregrounding may obscure.

We consider aesthetics as a way to surface these disciplinary blind spots. Here, *aesthetic perspectives* denote underlying assumptions about how and what is appreciated or overlooked (Gregg & Seigworth, 2020). Whilst IS research has considered interface design (Coursaris & van Osch, 2016; Tams & and Dulipovici, 2024), aesthetics has typically been reserved to instrumental concerns such as usability or managerial efficacy (Adam et al., 2024; Du et al., 2024; Guan et al., 2023). Attending to aesthetic perspectives reveals how perception and imagination are fundamental to understanding, from classical treatments of beauty (Plato, 2015) to contemporary affect theory (Gregg & Seigworth, 2020). Schopenhauer's account of aesthetic contemplation as temporary liberation from instrumental striving resonates here: in such moments, the self is freed from restless will and encounters reality beyond utility (Ciriello, 2025). Recent IS scholarship similarly calls for reflection on the philosophical assumptions that shape what counts as legitimate knowledge (Hassan et al., 2018; Lee & Sarker, 2023; Monteiro et al., 2022; Sullivan et al., 2024), yet aesthetic philosophy in these conversations remains largely unexplored. We build on this movement by foregrounding aesthetic perspectives as modes of knowing and valuing that matter for IS theory and practice. We ask: *What aesthetic categories and assumptions exist in IS research, and how might making these explicit expand the field's conceptual and normative horizons?*

We address this question through a hermeneutic literature analysis to uncover tacit assumptions shaping how IS phenomena are conceptualised (Boell, 2017; Boell & Cecez-Kecmanovic, 2014; Sullivan et al., 2024). Our analysis reveals distinct aesthetic perspectives shaping recognition, attention, and theorisation. We identify affect as a critical and undertheorised foundation, conceived either as measurable emotional reaction or pre-conscious atmospheric force shaping possibility (Gregg & Seigworth, 2020; Ingold, 2021). Building on this, we identify four aesthetic perspectives: aesthetics as *imitation* (representational fidelity and utility-aligned visual appeal); as *sensory experience* (embodied, pre-reflective, multi-sensory engagement); as *world-making* (imagination, attention, and sociotechnical imaginaries); and as *political doing* (aesthetic practices that reveal, contest, or reproduce power). These perspectives constitute foundational assumptions underpinning IS scholarship (Boell, 2017; Sullivan et al., 2024) yet remain largely implicit.

To illustrate their implications, we apply the perspectives to two cases: algorithmic management (an established IS concern) and digitally mediated intimacy (an emerging one). We show how different aesthetic perspectives open alternative research questions and highlight aspects that dominant framings may obscure. Algorithmic management is usually examined through accuracy, transparency, and fairness, whereas questions about how such systems feel, reshape professional identity, or shape perceptions receive less attention (Kellogg et al., 2020; Sullivan et al., 2024). Similarly, research on AI companions and dating platforms focuses on functionality or user satisfaction, rarely addressing connection, meaning, and embodiment (Chaturvedi et al., 2023; Wu & Kelly, 2020). These omissions suggest that what IS recognises as legitimate research concerns is shaped by aesthetic perspectives on what matters and whose experiences count (Sullivan et al., 2023).

Our contribution is threefold. First, we surface and synthesise implicit aesthetic assumptions into a conceptual framework that articulates how aesthetic perspectives shape theorising, method, and contribution. Second, we clarify the role of affect by distinguishing between two perspectives: affect as empirically measurable response and affect as atmospheric force that shapes what becomes perceptible. Third, through our two cases, we demonstrate how aesthetic analysis can be applied to established and emerging IS topics. Overall, we argue aesthetics is not peripheral to IS but constitutive of a humane, reflexive, and ethically ambitious discipline.

# 2 Background

This section reviews how aesthetics is described and deployed in IS research, highlighting variation in definitions and assumptions. Whilst such conceptual flexibility can be generative, it remains largely implicit, constraining the field's ability to harness aesthetics constructively.

## 2.1 Classical and Expressive Aesthetics

Aesthetics has long appeared in IS through design science (Baskerville et al., 2018), human-computer interactions (HCI) (Coursaris & van Osch, 2016), quantitative studies of enjoyment (Tams & and Dulipovici, 2024), qualitative work on identity (Hedman et al., 2019), and conceptual studies of appreciation in business processes (Blattmeier, 2023). Predominantly, aesthetics is treated as a binary measure of strict functionalism, defined as visual and hedonic qualities contrasted with usability and instrumental performance (Jiang, Weiquan, et al., 2016). Here, aesthetics is positioned as a measurable property to be controlled and optimised (Adam et al., 2024; Coursaris & van Osch, 2016), foregrounding how interface appearance shapes attention and judgement.

This view rests on the distinction between *classical aesthetics* that privilege clarity, order, and usability (e.g., Coursaris & van Osch, 2016) and *expressive aesthetics*, focused on creativity and novelty within design norms (Lavie & Tractinsky, 2004). Classical aesthetics presumes that symmetry and order enhance how users process information and interact with systems, whereas expressive aesthetics embraces unconventional design that breaks traditional norms but remain appealing to users. This framing extends into technology acceptance research where visual aesthetics and perceived enjoyment are positioned as antecedents of use intention and valued for predicting adoption over their intrinsic qualities (Van der Heijden, 2004).

## 2.2 Broader interpretations of Aesthetics in IS Research

Beyond instrumental variables, aesthetics also features in identity work and emotional relations with technology. Here, sensory and perceptual engagement shapes social identity (Bødker et al., 2014; Hedman et al., 2019). Users may value the affective and identity dimensions of technology over its functional outputs, as seen in the sense of belonging cultivated through membership in the 'iPhone club' (Bødker et al., 2014). Understanding technology use as embedded in the flow of everyday life rather than confined to discrete task-oriented episode can open up possibilities for aesthetic inquiry that moves beyond user behaviour (Yoo, 2010). Unlike HCI's focus on aesthetics as a design variable, this view treats aesthetics as inherent to relational engagement. It aligns with research showing how technology shapes attention "in-the-world", influencing sense-making, and identity (Hafermalz et al., 2020). Aesthetics thus becomes an inherent quality of human-technology relations. Recent work employs aesthetics for process theory. Blattmeier (2023) demonstrates how aesthetic modes of thinking and feeling reframe business processes beyond mechanistic metaphors, foregrounding atmospheres and affect as central to interpretation.

Within IS research, aesthetics has further served as a theoretical lens on sociotechnical life. Drawing on Foucault's aesthetics of resistance, Avgerou and McGrath (2007) show how aesthetic sensibility unsettles apolitical assumptions in technological rationality by situating it within power and practice. Foucault's (1978) 'aesthetics of existence' foregrounds life as an ongoing creative project, where moral and aesthetic forms mutually shape identity and social order. Instrumental rationality is thus not neutral but ethically and politically laden; surfacing aesthetic assumptions exposes the sociotechnical structures they sustain or contest. IS research leveraging Schopenhauer's philosophy similarly frames aesthetic experience as a suspension of striving: the world as "will" binds individuals to desire and suffering, whereas aesthetic contemplation enables perception freed from utility and egoistic interest. Such moments foster tranquillity, humility, and compassion, offering a non-instrumental mode of knowing that affirms shared humanity and resists optimisation-centred rationalities (Ciriello, 2025).

Overall, IS scholarship is widening its engagement with aesthetics, yet foundational assumptions remain largely implicit and weakly connected. Consequently, aesthetics appears in diffuse forms: as a design variable optimising usability and acceptance (Adam et al., 2024); as part of user experience, emphasising sensory engagement and emotion (Coursaris & van Osch, 2016; Tams & and Dulipovici, 2024); and as a source of identity and relational meaning, shaping how users define themselves and attend to technologies (Bødker et al., 2014; Hedman et al., 2019). Beyond experience, aesthetics functions as a mode of knowing and value orientation: enabling imagination and flourishing (Ciriello, 2025), informing theoretical critique (Avgerou & McGrath, 2007; Hassan et al., 2018; Sullivan, 2025), extending theory through attention to atmospheres and affect (Blattmeier, 2023) and fostering tacit, embodied understanding (Constantinides et al., 2012; Ingold, 2021). Yet these strands rarely engage one another or their diverse assumptions. This fragmentation hinders cumulative theorising and limits the field's ability to mobilise aesthetics as an epistemic resource.

# 3 Method: Hermeneutic Literature Analysis

Given the formative role of philosophical worldviews in shaping how IS phenomena are understood (Hassan et al., 2018) and calls to interrogate taken-for-granted assumptions (Monteiro et al., 2022), surfacing aesthetic commitments is essential for more reflective IS theorising. This is evident in emerging studies on digitally mediated intimacy, where platforms' curation of romantic encounters cannot be captured through usability metrics alone (Ciriello et al., 2024; Kurihara, 2025). User experiences of connection, desire, and meaning require attention to how systems *feel* in everyday life, what forms of relating they privilege, and whose emotional styles are foregrounded or sidelined.

To explore these issues, we conducted a hermeneutic literature analysis examining how aesthetics is understood in IS and how alternative aesthetic commitments might reshape future scholarship. Unlike systematic or critical reviews that synthesise existing knowledge (Paré et al., 2015), hermeneutic analysis foregrounds interpretation and surfaces the tacit assumptions that shape theorisation (Boell & Cecez-Kecmanovic, 2014), including their epistemic and normative consequences (Sullivan et al., 2024). Literature is treated as interpretive material, analysed iteratively to reveal implicit commitments and meaning structures. Our process followed three interconnected stages (sourcing, analysis, and assessment), though in practice these phases were cyclical and mutually informing to refine a typology of aesthetic assumptions in IS and their implications for theorising.

## 3.1 Sourcing

We began by identifying IS literature engaging aesthetics. Independent searches across the eleven AIS Senior Scholars' journals yielded 481 articles. These were screened to remove papers that used aesthetics only superficially or synonymously with interface appearance. Specifically, articles were included where aesthetics featured as a substantive theoretical or empirical concern, including as a design variable, a mode of knowing, or a site of political and affective engagement. Articles were excluded where the term appeared incidentally, as a synonym for visual appeal or surface design, without engagement with its conceptual implications. We then conducted forward and backward citation tracing (Boell, 2017; Sullivan et al., 2024), examining both sources referenced by key articles and subsequent work citing them. For example, (Avgerou & McGrath, 2007) discussion of rationality in IS led us to aesthetic orientation and motivated deeper engagement with philosophical aesthetics. This iterative tracing continued until theoretical saturation; that is, when additional articles yielded marginal new insights and converged on similar philosophical foundations, including Foucault (1977, 1978), Dewey (1934), and others. Finally, we consulted primary aesthetic philosophy to assess what IS foregrounds and overlooks, this revealed that IS scholarship engages extensively with philosophical perspectives (Hassan et al., 2018), aesthetic perspectives, such as those like Deleuze and Guattari (1987) who entwine aesthetics with ontology and epistemology, have received limited attention.

## 3.2 Analysis

The analysis proceeded through iterative framework development. We first drafted an initial schema based on patterns observed in IS and philosophical texts. Through collaborative refinement, we used *conceptual leaping* – moving between descriptive coding and abstract theorisation (Klag & Langley, 2013; Rivard, 2024) – to identify how aesthetic assumptions operate across the literature. This process crystallised four heuristic categories of aesthetic perspectives. Whilst acknowledging their overlap, we argue these categories offer analytic value for understanding aesthetic commitments in IS. To enhance interpretive rigour, we sought feedback from our peers in IS by presenting earlier iterations of our work (Boell, 2017; Sullivan et al., 2024). Where possible, we consulted authors whose work we analysed to validate interpretive accuracy, using scholarly dialogue to further refine and deepen our analysis.

Finally, echoing the methods of Sullivan et al., (2024), we treated the review process itself as part of our analysis, where our ideas were tested by the reviewers, leading to concrete changes in the presentation of our findings, such as the redevelopment of the map we present in the subsequent section and deepening how our perspective apply to the example cases we present.

Whilst replication of the sourcing process is possible, hermeneutic inquiry acknowledges that interpretation is perspectival for different readers and cannot be replicated, as they draw on different intellectual traditions, and arrive at different and equally valid categorisations (Boell & Cecez-Kecmanovic, 2014). The four perspectives we identify are outputs of our in-depth engagement with the literature, questioning our axiological assumptions in the IS research process. We acknowledge that our interpretive process was likely shaped by the philosophical orientations of our research environment, which foregrounds phenomenological and interpretivist traditions. This positionality potentially sensitised us to dimensions of aesthetic experience including the embodied, relational, and political aspects that may receive less attention in other researchers' readings of the same literature, as hermeneutic practice treats the researcher's horizon as a resource for interpretation rather than an obstacle to it (Boell, 2017). The contribution of this analysis therefore rests less on the reproducibility of the four categories per se, and more on the conceptual possibilities they reveal.

# 4 Findings: Surfacing Foundational Aesthetic Assumptions

Our hermeneutic analysis reveals that IS scholarship frames aesthetics in distinct ways, shaping how sociotechnical phenomena are perceived, valued, and theorised. We identify four foundational perspectives (*aesthetics as imitation*, as *sensory experience*, as *world-making*, and as *political doings*), each foregrounding appreciation and affect differently. These perspectives draw on seminal thinkers (e.g., Deleuze & Guattari, 1987; Heidegger, 1977; Kant & Goldthwait, 2003; Plato, 2015) and extend ontological, epistemological, and axiological commitments, shaping what becomes recognised and appreciated within IS research (Hassan et al., 2018; Sullivan et al., 2024). Before examining these perspectives, we first surface the affective foundations underpinning aesthetic assumptions, as affect emerged as a pervasive yet undertheorised dimension.

## 4.1 The Affective Foundations of Aesthetics

Affect underlies much aesthetic discourse but is situated in two fundamentally different perspectives. First, affect is treated as an underlying psychological temperament guiding attention (Deng & Poole, 2010). Second, and more radically, affect is understood as "visceral forces beneath, alongside, or generally other than conscious knowing" that "can drive us towards movement" or "suspend us" (Gregg & Seigworth, 2020). Thus, at the risk of oversimplification, we provisionally conceptualise *affect* as the preconscious forces shaping what becomes appreciable in experience, operating before language or deliberation, and what the "heart" registers precedes and potentially shapes what the "mind" is able to assess (Ciborra, 2006; Ingold, 2021; Merleau-Ponty, 1968). This distinction is particularly relevant to IS research as digital technologies increasingly mediate everyday experience. Understanding how aesthetic experiences unfold before conscious evaluation helps explain whether technologies feel inviting or alienating, meaningful or merely functional. These atmospheric forces materially shape interpretation and behaviour in IS contexts (Pignot & Thompson, 2024). We observe two dominant aesthetic orientations linked to affect: empirical aesthetics and phenomenological aesthetics.

First, *empirical aesthetics* concerns the visual design of systems. This view frames aesthetics in input–process–output terms, focusing on how design stimuli generate affective judgements and emotional responses (Konečni & Sargent-Pollock, 1977). Grounded in cognitivist assumptions (Ingold, 2021), affect here refers to a person's mood or subjective values which influencing how inputs are experienced. This view underpins IS work examining how artefacts influence users (e.g., Baskerville et al., 2018) or how aesthetics 'triggers' enjoyment (Tams & and Dulipovici, 2024). Empirical aesthetics supports UX and HCI research that quantifies aesthetic evaluation (e.g., Coursaris & van Osch, 2016), or examines it qualitatively (Du et al., 2024). Here, affect is treated as measurable, informing design guidelines and user experience models.

By contrast, *phenomenological aesthetics*, critiques input-process-output logics (e.g., Boland, 1985; Ingold, 2021). In this view, affect is not a discrete emotional state but a pre-conceptual, embodied force shaping conditions of possibility within a wider ecology of perception. Rather than measuring user reactions, this perspective explores how digital technologies create affective atmospheres and how users and technologies shape one another (Pedwell & Seigworth, 2023). Assuming aesthetics stems from embodied experiences grounded in relational ontology (Deleuze & Guattari, 1987; Heidegger, 1977), affect becomes a central aspect of *correspondence*, the process by which beings answer to one another through time (Ingold, 2021). Phenomenological aesthetics thus foregrounds flows of affectively charged experience over discrete cognitive episodes. Within IS, phenomenological aesthetic assumptions appear in studies embracing relational ontologies and process thinking (e.g., Baygi et al., 2021; Kreps & Rowe, 2024; Prester et al., 2023). For example, research on digital nomads shows how professional identities form through performative and sensory engagement with digital environments (Prester et al., 2023), treating affect and aesthetics as emerging through ongoing multisensory interaction.

In IS, empirical aesthetics supports UX models focused on appearance, usability, and satisfaction, treating appreciation as a discrete event. Phenomenological aesthetics foregrounds identity formation, spatiotemporal experience, and sense-making, framing appreciation as continuous embodied experiences. Both orientations underpin the four aesthetic perspectives we identify: *aesthetics as imitation* and *sensory experience* align with empirical affect, whilst *world-making* and *political doing* align with phenomenological affect, though these alignments overlap.

## 4.2 Aesthetic Perspectives in the IS Literature

Our analysis suggests that underlying divergences in IS scholarship on aesthetics can be clarified by attending to two analytical distinctions. The first distinction concerns the epistemic orientation of the aesthetic inquiry. *Object-oriented* perspectives locate aesthetic value in artefacts and their perceptible features, emphasising form, representation, and observable qualities. *Experience-oriented* perspectives locate aesthetics in lived experience, relational attunement, and embodied becoming, foregrounding how meaning unfolds through practice and perception. The second distinction concerns the nature of aesthetic knowledge. *Empirical* orientations view aesthetics as a discrete, measurable response to design, something that can be observed, tested, and verified independently of the experiencing subject. *Phenomenological* orientations view aesthetic experience as unfolding continuously through living it and inseparable from embodied encounter, atmosphere, and ongoing sensemaking.

Figure 1 organises these distinctions across two axes: the vertical axis distinguishes object-oriented from experience-oriented approaches, whilst the horizontal axis distinguishes empirical from phenomenological orientations, unlike some onto-epistemic commitments that may be philosophically irreconcilable (Riemer & Johnston, 2017), aesthetic orientations can overlap and interweave in productive ways. In mapping these perspectives we identified four distinct perspectives. First, *aesthetics as imitation* treats appreciation as recognising faithful or appealing representations of the material world. Second, *aesthetics as sensory experience* emphasises embodied, multisensory engagement with technologies, systems, and practices. Third, *aesthetics as world-making* highlights how patterns of (un)appreciation generate, sustain, and disrupt sociotechnical imaginaries. Fourth, *aesthetics as political doing* foregrounds aesthetic practices as performative acts that reinforce or transform power relations. Making these implicit assumptions explicit enables researchers to more deliberately choose aesthetic orientations suited to their questions and phenomena (Hassan et al., 2018; Sullivan et al., 2024). Note that these perspectives may overlap and combine in productive ways.

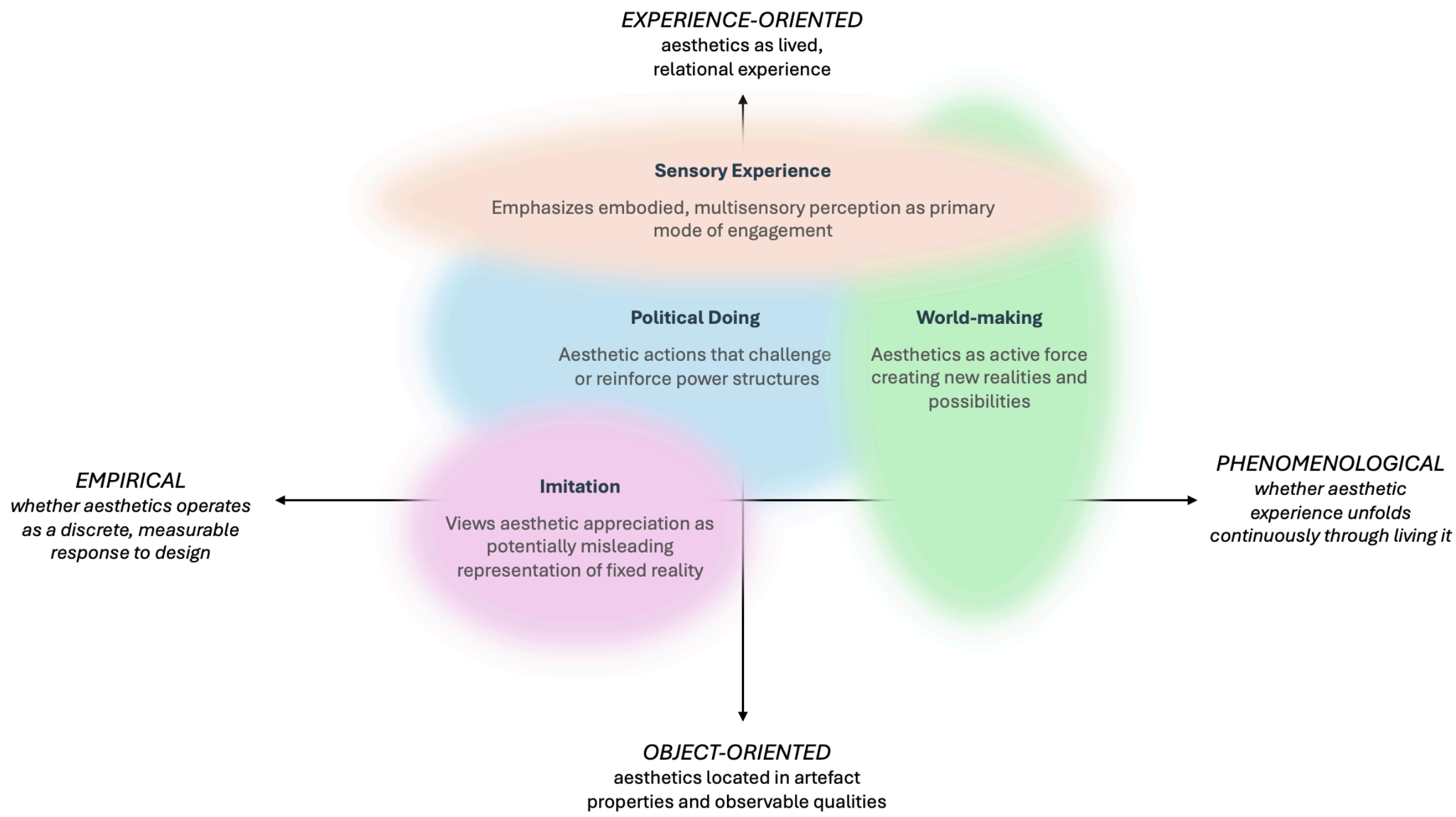

*Figure 1. Map of aesthetic perspectives across IS scholarship.*

### 4.2.1 Aesthetics as imitation

*Aesthetics as imitation* represents the most historically established perspective and underpins much empirical research. This view conceives aesthetics as the appreciation of representational fidelity, that is, how well an artefact or performance mirrors an external reality. This view matters because it positions beauty in relation to truth, raising questions about whether aesthetic delight clarifies or obscures it (Aristotle, 1997; Plato, 2015). Rooted in ancient Greek philosophy, this tradition troubles the relationship between appearance, knowledge, and moral character, a tension that also exists in IS scholarship's tendency to position aesthetics as a potentially distracting force requiring instrumental justification. Plato (2015) viewed art as deceptive imitation that seduces emotion and distances observers from truth and rationality whereas by contrast, Aristotle (1997) saw value in artistic imitation for offering emotional catharsis, though he still agreed aesthetics misrepresented and created an ontological distance to reality. Kierkegaard (1992) also focused on how aesthetic experiences could generate hedonic and morally corrosive behaviours. These foundational thinkers converge in their scepticism, noting the power of aesthetics to obscure truth and shape character in potentially undesirable ways. This helps to explain why IS research may have historically treated aesthetics with caution and confining it largely to controlled design variables.

In contemporary IS scholarship, this perspective appears in studies that treat aesthetics as means to stimulate engagement insofar as it advances functional or economic goals. Research on visual aesthetics and usability exemplifies this orientation: classical aesthetics (orderliness, clarity) and expressive aesthetics (creativity, novelty) are valued insofar as they improve satisfaction and system acceptance (Coursaris & van Osch, 2016; Tams & and Dulipovici, 2024). Similarly, e-commerce research has examined how aesthetic qualities influence trust and purchase intentions (Jiang, Wang, et al., 2016), operationalising aesthetics as a controllable design variable serving instrumental end. Standardised "beauty" or "visual appeal" scales in HCI (Adam et al., 2024) continue this lineage.

The underlying caution is that aesthetic experience, if unchecked, may obscure rational judgement, echoing Platonic suspicion that beauty and hedonic qualities requires management. In studies of digitally mediated intimacy, this perspective appears in evaluations of how convincingly AI companions simulate human connection through conversational design and personalisation (De Freitas et al., 2025). Here, aesthetic value lies in how well systems mirror "genuine" empathy (representational fidelity), treating emulated intimacy as a surface quality to be optimised whilst backgrounding ontological differences between human and artificial relating (Chen et al., 2023). Aesthetic appreciation is thus legitimised only when it enhances usability, efficiency, or acceptance, functioning as a controlled surface property whose value derives from instrumental ends.

#### 4.2.2 Aesthetics as sensory experience

*Aesthetics as sensory experience* foregrounds the immediate, embodied encounter between actors and their environments, emphasising how meaning arises through pre-reflective engagement before conscious interpretation (Merleau-Ponty, 1968). This perspective matters because it treats aesthetics not as a property of objects but as lived sensation: sight, sound, touch, taste, and scent shape what becomes salient and how affect circulates. This process creates *affective atmospheres*, referring to the felt intensities that generate action (Anderson, 2009). This view derives from phenomenology, where perception is reciprocal: actors do not merely perceive the world but participate in its construction (Merleau-Ponty, 1968), and it has direct implications for IS research as studying how users interact with systems cannot be separated from studying how those systems shape the conditions of perception and meaning-making itself.

This perspective has direct implications for IS because digital technologies increasingly mediate sensory experience. A foundational assumption in IS research holds that information can be encoded and transmitted through computational systems (Boell, 2017). Yet if aesthetics emerges through sensory experience, then what information is encoded, which modes of feeling are included or excluded, fundamentally shapes what can be appreciated. Consider algorithmic curation in social media, where recommendation systems privilege visual and textual modalities that induce intense emotions or cognitive disengagement. By selectively amplifying outrage, urgency, or distraction, these platforms shape not only what content users encounter but what forms of attention and appreciation become habitual. Virtual reality (VR) and augmented reality (AR) technologies reveal similar patterns of sensory privilege: typically vision and hearing whilst other sensory dimensions like smell, taste, ambient temperature remain largely absent (Wedel et al., 2020). This selective mediation opens questions about how sensory absences and presences together shape what becomes appreciable as "immersive" experience, and what forms of connection and embodied understanding become possible or constrained when entire sensory dimensions are backgrounded.

These sociotechnical developments raise normative questions: *whose* sensory world is encoded (or not), and *which* modes of feeling are included (or excluded)? For IS scholarship, this perspective highlights that design decisions do not merely influence usability. Whilst much IS research privileges visual and cognitive dimensions, sensory aesthetics points to opportunities for studying multisensory digital environments in domains like virtual education, aged care, and telehealth, where embodied engagement is central to human flourishing.

#### 4.2.3 Aesthetics as world-making

*Aesthetics as world-making* emerges from philosophical perspectives where worlds are not fixed backdrops but dynamic, ongoing creations shaped through ecological relations and collective imaginaries (Deleuze & Guattari, 1987). What matters here is not merely how things appear but how attention and appreciation bring particular realities into being, thereby co-creating what becomes thinkable, desirable, or actionable (de La Bellacasa, 2017; Ingold, 2021). This perspective assumes that aesthetics, as a process of attention and appreciation, informs what becomes real to actors. For instance, Heidegger (2010) described aesthetics as a mode of unconcealing, through which material substrates ("earth") acquire significance and worlds are disclosed. Deleuze and Guattari (1987) similarly frame aesthetics as a force of becoming that shapes identities and social structures, thus rendering them real.

In this way, *world-making* foregrounds the role of aesthetic sensibility in shaping *sociotechnical imaginaries*: collective visions of what technologies are and might become (Jasanoff, 2020). When aesthetic perspectives guide what is valued and noticed, certain futures appear plausible whilst others recede, structuring what is experienced as real or possible. This has direct implications for IS, where digital platforms and algorithmic systems increasingly curate attention, direct imagination, and scaffold expectations (Hovorka et al., 2025; Kurihara, 2025). Search engines, recommender systems, and interface cues do not simply reflect worlds; they condition what users see, feel, and consider meaningful, thereby influencing identity, aspiration, and action.

For IS scholarship, this perspective highlights how digital infrastructures participate in world-building. It offers an analytical lens for examining how sociotechnical systems cultivate imaginaries, reconfigure professions, and reshape conditions of possibility (Baygi et al., 2024; Hovorka & Mueller, 2025; Márton, 2021).

### 4.2.4 Aesthetics as political doings

The fourth perspective, *aesthetics as political doings*, views aesthetic experience as a site of struggle where perception, meaning, and possibility are contested and reconfigured. What matters here is not just what is sensed, but how sensory and affective arrangements shape visibility, legitimacy, and action. It overlaps with sensory and world-making perspectives, yet places explicit emphasis on intention, critique, and political consequence. Here, aesthetics becomes a site where political ideals are enacted or constrained through shifts in visibility, intelligibility, and affect (Susskind, 2018).

This view draws on traditions that cast aesthetics as a force of disruption and emancipation. Thinkers such as Adorno, Foucault, Hegel, Nietzsche, Rancière, and Schopenhauer treat aesthetics as capable of unsettling hegemonic forms of life and enabling new ways of being and knowing. Rancière and Rockhill (2012) argue that aesthetics is inherently political because it structures the ‘distribution of the sensible’, that is, what is perceptible, valued, or ignored. Foucault (1978) concept of the aesthetics of existence similarly frames self-fashioning as a political practice through which subjects comply with or resist domination. In algorithmic management and digitally mediated intimacy, such aesthetic arrangements shape whose labour or loneliness becomes visible, whose experiences are validated, and what futures appear inevitable rather than contestable.

In IS scholarship, this perspective can help unpack how digital technologies encode and distribute aesthetic values that carry political force. Aesthetic arrangements in interfaces, data visualisations, and algorithmic curation shape who is seen, whose voices are amplified or suppressed, and what norms are reinforced. Recent work extends this view, examining how aesthetic values underpin systems of governance and resistance (Avgerou & McGrath, 2007; Ciriello, 2025). For instance, Benfeldt et al. (2025) show how a Nordic aesthetic of data governance, centred on minimalism and collective responsibility, reconfigures sociopolitical relations and institutional norms. In sum, aesthetics as political doing invites scholars to interrogate how aesthetics reinforce or challenge power, and how alternative aesthetic imaginaries might support more equitable, humane, and plural futures.

# 5 Discussion

This hermeneutic analysis surfaces and synthesises aesthetic perspectives in IS scholarship, demonstrating that aesthetics shapes the assumptions IS researchers bring to their work, and determine what becomes recognisable as legitimate research, how phenomena are theorised, and what forms of flourishing appear possible in sociotechnical systems. We do not propose aesthetic inquiry as a replacement for instrumental concerns, but as their reflexive complement. Instrumental rationality itself carries aesthetic commitment, about clarity, efficiency, and measurability that privilege certain experiences whilst backgrounding others. Our purpose is to expose this partiality and extend the epistemic range of the field.

## 5.1 Implications of aesthetic perspectives for IS scholarship

Aesthetic perspectives influence how established IS phenomena are made sense of, with commitments regarding appreciation shaping what is emphasised or overlooked. Rather than asking whether technologies are effective, usable, or efficient at a given moment, these perspectives ask how systems *come to be appreciated*, how felt experience evolves, and how technologies shape meaning. Although these perspectives overlap, articulating their assumptions remains vital: what a field deems aesthetically legitimate shapes what it recognises as theoretically legitimate (Ingold, 2021). Thus appreciation becomes processual rather than episodic; affect becomes foundational to sense-making (Gregg & Seigworth, 2020). When aesthetics is confined to usability or efficiency, IS research privileges calculative rationality and sidelines embodied, affective, and imaginative dimensions of sociotechnical life. When approached as relational attunement or world-making, different questions arise.

We illustrate this reorientation through two examples, algorithmic management and digitally mediated intimacy. In doing so, we show how the four aesthetic perspectives reveal otherwise obscured aspects of IS phenomena. The first, algorithmic management, concerns systems that monitor, evaluate, and govern work (Kellogg et al., 2020; Sullivan et al., 2024). The second, digitally mediated intimacy, such as AI companions, artificial emotional intelligence (AEI), and dating platforms, shows how digital systems curate emotional life and reshape the conditions of human connection (Ciriello et al., 2024; Kurihara, 2025). Applying our framework reveals how aesthetics as *sensory experience*, aesthetics as *world-making* , and aesthetics as *political doing* generate fundamentally different research questions and expose dimensions that dominant framings systematically obscure. These illustrative examples were chosen on the basis that aesthetics plays a role in understanding established phenomena of inquiry (algorithmic management) and emerging research concerns (digitally mediate intimacy).

| Perspective | What the perspective surfaces | Questions enabled |
|---|---|---|
| *Aesthetics as Imitation* | Foregrounds: algorithmic accuracy, representational fidelity, system neutrality; shaping what IS research treats as a legitimate performance metric and what counts as organisational truth<br>Backgrounds: the normative and political work done by institutional processes and embedded biases that polished interface design may conceal | *How do aesthetically appealing interfaces shape managerial trust in algorithm-produced insights?*<br>*What institutional biases are masked by interface designs that present system recommendations as neutral or purely data-driven?* |
| *Aesthetics as Sensory Experience* | Foregrounds: the immediate, embodied encounter between workers and algorithmic systems; directing IS researchers toward how app design and gamification shape what management feels like before workers consciously evaluate it<br>Backgrounds: the assumption that abstract utility is a sufficient basis for evaluating algorithmic work systems | *How do interface design choices create affective atmospheres for workers? How does the embodied experience of algorithmic management shape professional identity and emotional wellbeing?* |
| *Aesthetics as World-Making* | Foregrounds: how design choices and visualisations actively construct organisational realities, i.e.. performance metrics and visualisation choices do not merely represent work but actively constitute what 'good performance' means, which professional identities become intelligible, and what forms of labour are recognised as valuable<br>Backgrounds: Algorithmic management as neutral coordination | *What alternative visions of work become unimaginable through design choices?*<br>*How do performance metrics reshape the meaning of professionalism, authority, and organisational value?* |
| *Aesthetics as Political Doing* | Foregrounds: How aesthetic arrangements in dashboards and interfaces encode and reproduce power, directing IS research toward whose labour is made visible, whose is obscured, and what ideologies are naturalised through design<br>Backgrounds: Algorithmic management as technical optimisation, which forecloses critical inquiry into its distributional and political consequences | *How do performance dashboards function as instruments reinforcing particular ideologies?*<br>*What alternative aesthetic frameworks might support worker power, equity, and collective organising?* |

*Table 1. Aesthetic perspectives applied to Algorithmic Management*

Table 1 summarises what each aesthetic perceptive foregrounds and backgrounds in **algorithmic management**. Research here has been dominated by *aesthetics as imitation*, privileging accuracy and transparency whilst overlooking embodied and political dimensions.

*Aesthetics as imitation* foregrounds algorithmic accuracy and representational fidelity, evaluating how faithfully systems mirror organizational rationality through clean interfaces and transparent logic (Lamers et al., 2024). Industrial aesthetics (efficiency, productivity, datafication) are appreciated, positioning system outputs as neutral or objectively correct (Kordzadeh & Ghasemaghaei, 2022). This perspective backgrounds the underlying institutional processes and implicit biases embedded in models, treating algorithms as value-neutral representations rather than normatively laden constructions. The risk is that representational fidelity becomes a surface quality overlooking more deeply sociotechnical aspects of algorithmic management.

*Aesthetics as sensory experience* foregrounds the immediate, embodied encounter between workers and algorithmic systems, how app design, algorithmic nudging, audio cues, and gamification shape what management *feels like*. Pre-conscious sensory dimensions (enjoyment, friction, exhaustion) guide engagement, compliance, and resistance before cognitive evaluation (Kellogg et al., 2020). This perspective backgrounds abstract utility and measurable outcomes, treating algorithmic management not as information delivery but as affective mediation that conditions identity and belonging in algorithmically governed work (Schildt, 2016).

*Aesthetics as world-making* foregrounds how algorithmic systems actively construct organizational realities through attention and appreciation. Design choices, metrics, and visualizations guide what becomes conceivable, desirable, and actionable; what is measured becomes what matters. Algorithms disclose worlds (e.g., dashboards foregrounding productivity reshape what a “worker” is) whilst obscuring alternative possibilities. This perspective backgrounds algorithmic management as neutral coordination, instead treating it as generative infrastructure that scaffolds organizational imaginaries and conditions future possibility.

*Aesthetics as political doing* foregrounds how aesthetic arrangements encode and reproduce power relations by structuring what is perceptible, valued, or overlooked. Algorithms curate visibility, amplify certain voices, and constrain political possibilities through what is datafied, disclosed, or obscured. Performance dashboards function as instruments of governance, reinforcing managerial ideology by privileging industrial aesthetics (efficiency, optimisation) (Lamers et al., 2024). This perspective backgrounds algorithmic management as technical optimisation, instead treating interface design and algorithmic filtering as political acts that shape patterns of inclusion, exclusion, and recognition.

Table 2 extends the analysis to **digitally mediated intimacy**. Here, aesthetic choices shape whose emotional lives are valued and how authenticity is defined. By contrasting perspectives, we show how IS scholars might move beyond evaluating realism or engagement to examining how digitally mediated intimacy systems configure desire, loneliness, and recognition.

*Aesthetics as imitation* foregrounds how realistically platforms simulate human empathy, courtship, and connection through conversational design and emulated responsiveness. AI companions are evaluated by representational fidelity, that is, how convincingly they mirror ‘genuine’ human interaction through textual depth, emotional range, and personalized responses (Chen et al., 2023). This perspective backgrounds the fundamentally different ontology of artificial relating, treating emulated empathy as a surface quality that can be optimised for convincingness. The risk is that improved simulation raises ethical concerns about deception whilst appearing ethically permissible through technical achievement.

*Aesthetics as sensory experience* foregrounds the sensory dimensions of user interaction with intimate technologies, that is, how interfaces mediate embodied encounters through visual, auditory, and haptic channels. The immediate, felt experience in AI companionship reveals sensory absences like physical warmth, reciprocal presence, and multisensory connection that characterize face-to-face intimacy. Visual design of dating apps privileges particular forms of attraction (images, text descriptions) whilst backgrounding sensory modes (touch, smell, ambient co-presence) that traditionally ground intimacy. This perspective backgrounds abstract utility and matching algorithms, treating platforms as affective mediators that condition what forms of sensory connection become possible or impossible, and how desire and attachment are felt.

| Perspective | What the perspective surfaces | Questions enabled |
|---|---|---|
| *Aesthetics as Imitation* | Foregrounds: how realistically platforms simulate human empathy, courtship, and connection; orienting IS research toward optimising conversational design and personalisation as proxies for genuine human connection | *How can we design more convincing simulations of human empathy and companionship?*<br>*At what point does representational fidelity become deceptive?* |
| | Backgrounds: the fundamentally different ontological status of artificial relating, and the ethical implications of treating emulated intimacy as a surface quality to be refined | |
| *Aesthetics as Sensory Experience* | Foregrounds: Sensory dimensions of embodied interaction (visual, auditory, haptic); revealing what is structurally absent from digitally mediated intimacy and directing IS research toward how sensory constraints shape what forms of connection become possible or foreclosed | *How do sensory limitations shape what feels 'real' or 'sustaining' in AI companionship?*<br>*How does the absence of reciprocal touch, smell, or spatial co-presence reshape the experience of romantic encounters when it is digital?* |
| | Backgrounds: the assumption that abstract utility and matching algorithms are adequate measures of relational quality | |
| *Aesthetics as World-Making* | Foregrounds: How platforms shape sociotechnical imaginaries of intimacy, conditioning collective expectations about desirability, commitment, and what constitutes successful connection, thus direct implications for how IS researchers frame and evaluate platform design | *What forms of relating become (un)imaginable through dominant design patterns (e.g., fast courtship, ambiguous connection)?*<br>*What alternative imaginaries of companionship might technology enable beyond optimisation and efficiency?* |
| | Backgrounds: technical functionality and satisfaction metrics, which forecloses inquiry into how platforms reshape the meaning of human connection | |
| *Aesthetics as Political Doing* | Foregrounds: Whose loneliness is commodified, whose intimacy is validated, and what norms of gender, desirability, and emotional labour are encoded in interface design. | *How do design choices reinforce or challenge norms around gender, desirability, and emotional labour?*<br>*What alternative aesthetic frameworks might resist commercialisation of loneliness and support collective care and solidarity?* |
| | Backgrounds: Technical design and user experience, which obscures the political and commercial dimensions of digitally mediated intimacy | |

*Table 2. Aesthetic perspectives applied to Digitally Mediated Intimacy*

*Aesthetics as world-making* foregrounds how dating platforms and AI companions shape sociotechnical imaginaries of intimacy, structuring expectations about relationship norms, desirability criteria, and what constitutes 'successful' connection. Design choices (swipe mechanics, compatibility algorithms, conversational scripts, which personal information to share) guide what becomes thinkable and desirable in relating, conditioning collective meanings of partnership, loneliness, and commitment (Kurihara, 2025). This perspective backgrounds technical functionality and user satisfaction metrics, focusing instead on how platforms reshape collective imagination.

*Aesthetics as political doing* perspective foregrounds how AI companionship and dating platforms encode aesthetic values that carry political force. Design choices structure whose loneliness is commodified, whose intimacy is deemed worthy of human investment, and what forms of relating are validated or pathologized through interface norms and commercial extraction. Platforms operate as aesthetic regimes that distribute recognition, desire, and belonging unequally. This perspective backgrounds technical design and user experience, focusing instead on how platforms function as political instruments that encode power relations.

Collectively, these perspectives invite a reconsideration of dominant approach to IS theorising, shifting attention towards processual and situated nature of aesthetics in sociotechnical phenomena. Rather than treating aesthetics as a secondary concern, we suggest that what is perceived, felt and appreciated plays a primary role in shaping the contours how IS phenomena are experienced. In this view, appreciation is not merely a response to design but a constitutive element in how sociotechnical phenomena is interpreted and understood in IS research.

## 5.2 Limitations and Future Research Direction

Our inquiry is an initial exploration of what aesthetic philosophy can offer IS scholarship, rather than a definitive account. Dominant IS perspectives, shaped largely by Western traditions valuing control, symmetry, seamlessness, and instrumental clarity have supported major advances but can also limit imagination and ethical considerations (Chughtai & Young, 2024). Aesthetic pluralism encourages sensitivity to alternative values and modes of being. Many non-Western aesthetic traditions remain unexplored in IS. For instance, the Japanese aesthetic philosophy of Wabi-sabi, a Japanese aesthetic philosophy embraces impermanence, incompleteness, and the beauty of wear and repair (Juniper, 2013). Applied to IS, it suggests systems that reveal their histories of use, evolve openly, and honour uncertainty rather than masking it with polished finality. Wikipedia’s visible edit trails and participatory civic platforms reflect this ethos of systems as evolving commons. Ubuntu aesthetics offers another example, emphasising relational interdependence, ‘I am because we are’ (Abubakre et al., 2021). Such a perspective on non-Western aesthetics was beyond our analysis due to their absence in the IS body of literature, yet such perspective raise questions of what IS technologies would be like from a different aesthetic origin.

These traditions are not decorative add-ons but ontological alternatives. They illustrate that dominant values (such as seamlessness, speed, optimisation, individual autonomy) are aesthetic commitments, not universal truths. A pluralist stance encourages IS researchers to ask: whose aesthetics shape system design? Whose forms of flourishing are centred or neglected? What might technologies look like if they were designed for resilience, interdependence, or cyclical renewal rather than perpetual acceleration? Methodologically, aesthetic pluralism encourages sensory ethnography, phenomenological interviewing, atmospheric analysis, design anthropology, and participatory aesthetic prototyping. It also calls for comparative aesthetic studies across cultural contexts, resisting universalising design principles and cultivating sensitivity to diverse perceptual and moral ecologies.

# 6 Conclusion

This paper argues that aesthetics is not peripheral to IS research, but constitutive to how the field recognises, values, and theorises sociotechnical phenomena. Through a hermeneutic literature analysis, we surfaced four aesthetic perspectives already operating, often implicitly, across IS scholarship. These perspectives shape what researchers attend to, what forms of experience become thinkable, and what kinds of sociotechnical futures become desirable or actionable. Our central claim is not that IS should replace instrumental concerns with aesthetic ones, but that instrumental rationality is itself just one among several legitimate aesthetic commitments of IS scholarship. When efficiency, clarity, and optimisation dominate as default criteria of relevance, embodied experience, affective atmospheres, imagination, and political visibility can recede from view. Making diverse aesthetic perspectives explicit broadens the field’s conceptual horizon and allows familiar phenomena to be seen not only as systems to be evaluated, but as arrangements that shape perception, meaning, and human flourishing.

If, as Schopenhauer reminds us in opening this paper, aesthetic contemplation offers temporary release from the suffering bound up with instrumental striving, then attending to aesthetics in IS research can offer more than prettier interfaces. It becomes a way of questioning what our dominant forms of inquiry make visible, what they overlook, and what digital ways of living they privilege or marginalise. Attending to aesthetics, then, ultimately means *paying attention to what we pay attention to.*

## References

Abubakre, M., Faik, I., & Mkansi, M. (2021). Digital Entrepreneurship and Indigenous Value Systems: An Ubuntu Perspective. *Information Systems Journal*, *31*(6), 838-862. https://doi.org/10.1111/isj.12343

Adam, M., Lins, S., Sunyaev, A., & Benlian, A. (2024). The contingent effects of is certifications on the trustworthiness of websites. *Journal of the Association for Information Systems*, *25*(3), 549-617. https://doi.org/10.17705/1jais.00836

Anderson, B. (2009). Affective Atmospheres. *Emotion, Space and Society*, *2*, 77-81. https://doi.org/10.1016/j.emospa.2009.08.005

Aristotle. (1997). *The poetics*. Penguin Classics.

Avgerou, C., & Mcgrath, K. (2007). Power, Rationality, and the Art of Living through Socio-Technical Change. *MIS Quarterly*, *31*(2), 295-315. https://doi.org/10.2307/25148792

Baskerville, R. L., Mala, K., & And Storey, V. C. (2018). Aesthetics in design science research. *European Journal of Information Systems*, *27*(2), 140-153. https://doi.org/10.1080/0960085X.2017.1395545

Baygi, R. M., Introna, L. D., & Hultin, L. (2021). Everything flows: Studying continuous sociotechnological transformation in a fluid and dynamic digital world. *MIS Quarterly*, *45*(1), 423-452. https://doi.org/10.25300/MISQ/2021/15887

Baygi, R. M., Introna, L. D., & Ostovar, M. (2024). Beyond Categories: A Flow-Oriented Approach To Social Justice on Online Labor Platforms1. *Management Information Systems Quarterly*, *48*(4), 1663-1690. https://doi.org/10.25300/MISQ/2024/18253

Benfeldt, O., Schroder, A., Zambach, S., Greve, M., Singh, R., & Gierlich-Joas, M. (2025). From context to aesthetic: A Nordic perspective on data studies in information systems.

Blattmeier, M. (2023). The aestheticization of business processes: Visualizing their Gestalt for collective thinking. *Journal of Information Technology*, *38*(4), 459-486. https://doi.org/10.1177/02683962231166438

Bødker, M., Gimpel, G., & Hedman, J. (2014). Time-out/time-in: the dynamics of everyday experiential computing devices. *Information Systems Journal*, *24*(2), 143-166. https://doi.org/10.1111/isj.12002

Boell, S. K. (2017). Information: Fundamental positions and their implications for information systems research, education and practice. *Information and Organization*, *27*(1), 1-16. https://doi.org/10.1016/j.infoandorg.2016.11.002

Boell, S. K., & Cecez-Kecmanovic, D. (2014). A Hermeneutic Approach for Conducting Literature Reviews and Literature Searches. *Communications of the Association for Information Systems*, *34*(12). https://doi.org/10.17705/1CAIS.03412

Boland, R. J. (1985). Phenomenology: A Preferred Approach to Research on Information Systems. *Research Methods in Information Systems*, 193-201.

Chaturvedi, R., Verma, S., Das, R., & Dwivedi, Y. K. (2023). Social companionship with artificial intelligence: Recent trends and future avenues. *Technological Forecasting and Social Change*, *193*, 122634. https://doi.org/10.1016/j.techfore.2023.122634

Chen, A., Kögel, S., Hannon, O., & Ciriello, R. F. (2023). *Feels Like Empathy: How "Emotional" AI Challenges Human Essence* Australasian Conference on Information Systems (ACIS2023), Wellington, New Zealand.

Chughtai, H., & Young, A. G. (2024). Decoloniality and Information Systems: Making Local Contexts Relevant to Is Research. *Information Systems Journal*. https://doi.org/10.1111/isj.12579

Ciborra, C. (2006). The Mind or the Heart? It Depends on the (Definition of) Situation. *Journal of Information Technology*, *21*(3), 129-139. https://doi.org/10.1057/palgrave.jit.2000062

Ciriello, R. (2025). Dear Schopenhauer: How can IS artefacts help us flourish? Wille, Vorstellung, compassion, and beauty in social media. *Scandinavian Journal of Information Systems*, *37*(1), 317-354. https://doi.org/10.17705/3SJIS/037.09

Ciriello, R. F., Hannon, O., Chen, A., & Vaast, E. (2024). Ethical Tensions in Human-AI Companionship: A Dialectical Inquiry into Replika. *Hawaii International Conference on System Sciences (HICSS2024).*

Clarke, R., & Davison, R. M. (2020). Research perspectives: Through whose eyes? The critical concept of researcher perspective. *Journal of the Association for Information Systems*, *21*(2), 483-501. https://doi.org/10.17705/1jais.00609

Constantinides, P., Chiasson, M., & Introna, L. (2012). The Ends of Information Systems Research: A Pragmatic Framework. *MIS Quarterly*, *36*, 1-20. https://doi.org/10.2307/41410403

Coursaris, C. K., & Van Osch, W. (2016). A Cognitive-Affective Model of Perceived User Satisfaction (CAMPUS): The complementary effects and interdependence of usability and aesthetics in IS design. *Information & Management*, *53*(2), 252-264. https://doi.org/10.1016/j.im.2015.10.003

De Freitas, J., Uğuralp, Z. O., Uğuralp, A. K., & Puntoni, S. (2025). AI Companions Reduce Loneliness. *Journal of Consumer Research*. https://doi.org/10.1093/jcr/ucaf040

De La Bellacasa, M. P. (2017). *Matters of care: Speculative ethics in more than human worlds* (Vol. 41). U of Minnesota Press.

Deleuze, G., & Guattari, F. (1987). *A Thousand Plateaus: Capitalism and Schizophrenia (B. Massumi, Trans.).* . University of Minnesota Press. (Original work published 1980).

Deng, L., & Poole, M. (2010). Affect in Web Interfaces: A Study of the Impacts of Web Page Visual Complexity and Order. *MIS Quarterly*, *34*, 711-730. https://doi.org/10.2307/25750702

Dewey, J. (1934). *Art as Experience*. Balch & Company.

Du, W., Ghorbani, M., Ni, Z., & Pan, S. L. (2024). Sustainable affordances of information systems for cultural tourism: An organisational aesthetics perspective. *Information Systems Journal*, *34*(5), 1787-1809. https://doi.org/10.1111/isj.12498

Foucault, M. (1977). *Discipline and Punish: The Birth of the Prison*. Penguin.

Foucault, M. (1978). *The Will to Knowledge: The History of Sexuality Vol. I*. Penguin Books.

Gregg, M., & Seigworth, G. J. (2020). *The affect theory reader*. Duke University Press.

Guan, Y., Tan, Y., Wei, Q., & Chen, G. (2023). When images backfire: The effect of customer-generated images on product rating dynamics. *Information Systems Research*, *34*(4), 1641-1663. https://doi.org/10.2139/ssrn.3633590

Hafermalz, E., Johnston, R. B., Hovorka, D. S., & Riemer, K. (2020). Beyond 'mobility': A new understanding of moving with technology. *Information Systems Journal*, *30*(4), 762-786. https://doi.org/10.1111/isj.12283

Hassan, N. R., Mingers, J., & Stahl, B. (2018). Philosophy and information systems: where are we and where should we go? *European Journal of Information Systems*, *27*(3), 263-277. https://doi.org/10.1080/0960085X.2018.1470776

Hedman, J., Bødker, M., Gimpel, G., & Damsgaard, J. (2019). Translating evolving technology use into user stories: Technology life narratives of consumer technology use. *Information Systems Journal*, *29*(6), 1178-1200. https://doi.org/10.1111/isj.12232

Heidegger, M. (1977). *The question concerning technology and other essays* (Paperback ed.). Harper & Row.

Heidegger, M. (2010). *Being and time*. SUNY press.

Hovorka, D., Thoring, K., & Mueller, B. (2025). *Pushing the Boundaries of Reality: Imagination and Design in Science*. https://doi.org/10.24251/HICSS.2025.730

Hovorka, D. S., & Mueller, B. (2025). Speculative foresight: A foray beyond digital transformation. *Information Systems Journal*, *35*(1), 140-162. https://doi.org/10.1111/isj.12530

Ingold, T. (2021). *Imagining for real: Essays on creation, attention and correspondence*. Routledge. https://doi.org/10.4324/9781003110910

Jasanoff, S. (2020). Imagined worlds: The politics of future-making in the twenty-first century. In *The politics and science of prevision* (pp. 27-44). Routledge. https://doi.org/10.4324/9781003022428-4

Jiang, Z., Wang, W., Tan, B., & Yu, J. (2016). The Determinants and Impacts of Aesthetics in Users' First Interaction with Websites. *Journal of Management Information Systems*, *33*, 229-259. https://doi.org/10.1080/07421222.2016.1172443

Jiang, Z., Weiquan, W., Y., T. B. C., & And Yu, J. (2016). The Determinants and Impacts of Aesthetics in Users' First Interaction with Websites. *Journal of Management Information Systems*, *33*(1), 229-259. https://doi.org/10.1080/07421222.2016.1172443

Juniper, A. (2013). *Wabi Sabi: The Japanese Art of Impermanence*. Tuttle Publishing.

Kant, I., & Goldthwait, J. T. (2003). *Observations on the feeling of the beautiful and sublime* (2nd paperback ed.). University of California Press.

Kellogg, K. C., Valentine, M. A., & Christin, A. (2020). Algorithms at work: The new contested terrain of control. *Academy of Management Annals*, *14*(1), 366-410. https://doi.org/10.5465/annals.2018.0174

Kierkegaard, S. (1992). *The Concept of Anxiety: A Simple Psychologically Oriented Deliberation in View of the Dogmatic Problem of Hereditary Sin*. Princeton University Press.

Klag, M., & Langley, A. (2013). Approaching the Conceptual Leap in Qualitative Research. *International Journal of Management Reviews*, *15*(2), 149-166. https://doi.org/10.1111/j.1468-2370.2012.00349.x

Konečni, V. J., & Sargent-Pollock, D. (1977). Arousal, positive and negative affect, and preference for Renaissance and 20th-century paintings. *Motivation and Emotion*, *1*(1), 75-93. https://doi.org/10.1007/BF00997582

Kordzadeh, N., & Ghasemaghaei, M. (2022). Algorithmic bias: review, synthesis, and future research directions. *European Journal of Information Systems*, *31*(3), 388-409. https://doi.org/10.1080/0960085X.2021.1927212

Kreps, D., & Rowe, F. (2024). Research Perspectives: Using a Process Philosophy Perspective in Information Systems Research: Principles of Creative Evolution. *Journal of the Association for Information Systems*, *25*(5), 1410-1433. https://doi.org/10.17705/1jais.00883

Kudina, O. (2021). "Alexa, who am I?": Voice assistants and hermeneutic lemniscate as the technologically mediated sense-making. *Human Studies*, *44*(2), 233-253. https://doi.org/10.1007/s10746-021-09572-9

Kurihara, F. (2025). What are you swiping for? Exploring user assessments of algorithms-based recommendations on online dating platforms. European Conference on Information Systems (ECIS 2025), Amman, Jordan.

Lamers, L., Meijerink, J., & Rettagliata, G. (2024). Blinded by "algo economicus": Reflecting on the assumptions of algorithmic management research to move forward. *Human Resource Management*, *63*(3), 413-426. https://doi.org/10.1002/hrm.22204

Lavie, T., & Tractinsky, N. (2004). Assessing dimensions of perceived visual aesthetics of web sites. *International Journal of Human-Computer Studies*, *60*(3), 269-298. https://doi.org/10.1016/j.ijhcs.2003.09.002

Lee, A. S., & Sarker, S. (2023). A Philosophical Perspective on Qualitative Research in the Age of Digitalization. In B. Simeonova & R. D. Galliers (Eds.), *Cambridge Handbook of Qualitative Digital Research* (pp. 15-27). Cambridge University Press. https://doi.org/10.1017/9781009106436.004

Márton, A. (2021). Steps toward a digital ecology: ecological principles for the study of digital ecosystems. *Journal of Information Technology*, *37*(3), 250-265. https://doi.org/10.1177/02683962211043222

Merleau-Ponty, M. (1968). *The Visible and the Invisible: followed by working notes*. Northwestern University Press.

Monteiro, E., Constantinides, P., Scott, S., Shaikh, M., & Burton-Jones, A. (2022). Editor's comments: qualitative research methods in information systems: a call for phenomenonfocused problematization. *MIS Quarterly*, *46*(4), iii-xix. https://doi.org/10.25300/MISQ/2022/464E1

Paré, G., Trudel, M.-C., Jaana, M., & Kitsiou, S. (2015). Synthesizing information systems knowledge: A typology of literature reviews. *Information & Management*, *52*(2), 183-199. https://doi.org/10.1016/j.im.2014.08.008

Pariser, E. (2011). *The filter bubble: What the Internet is hiding from you*. Penguin Press.

Pedwell, C., & Seigworth, G. (2023). The Affect Theory Reader 2 INTRODUCTION. In. Duke University Press.

Pignot, E., & Thompson, M. (2024). Affect and relational agency: How a negative ontology can broaden our understanding of IS research. *Information and Organization*, *34*(1), 100500. https://doi.org/10.1016/j.infoandorg.2023.100500

Plato. (2015). *The Republic*. Lerner Publishing Group.

Prester, J., Cecez-Kecmanovic, D., & Schlagwein, D. (2023). Toward a theory of identity performance in unsettled digital work: The becoming of 'digital nomads'. *Journal of Information Technology*, *38*(4), 442-458. https://doi.org/10.1177/02683962231196310

Rancière, J., & Rockhill, G. (2012). *The Politics of Aesthetics: The Distribution of the Sensible* ([New] edition. ed.). Bloomsbury Publishing Plc. https://doi.org/10.5040/9781350284913

Riemer, K., & Johnston, R. B. (2017). Clarifying Ontological Inseparability with Heidegger's Analysis of Equipment. *MIS Quarterly*, *41*(4), 1059-1082. https://doi.org/10.25300/MISQ/2017/41.4.03

Riemer, K., & Peter, S. (2021). Algorithmic audiencing: Why we need to rethink free speech on social media. *Journal of Information Technology*, *36*(4), 409-426. https://doi.org/10.1177/02683962211013358

Rivard, S. (2024). Unpacking the process of conceptual leaping in the conduct of literature reviews. *The Journal of Strategic Information Systems*, *33*(1), 101822. https://doi.org/10.1016/j.jsis.2024.101822

Schildt, H. (2016). Big data and organizational design – the brave new world of algorithmic management and computer augmented transparency. *Innovation*, *19*(1), 23-30. https://doi.org/10.1080/14479338.2016.1252043

Sullivan, R. (2025). Vulnerability, Poetics, and Mysticism: Troubling the taboo of unsanitized writing in Information Systems. *Communications of the Association for Information Systems*, *57*(1), 23. https://doi.org/10.17705/1CAIS.05723

Sullivan, R., Hannon, O., & Hovorka, D. (2023). Dark Matters: Becoming Attuned to Epistemic Darkness. ACIS 2023 Proceedings,

Sullivan, R., Veen, A., & Riemer, K. (2024). Furthering engaged algorithmic management research: Surfacing foundational positions through a hermeneutic literature analysis. *Information and Organization*, *34*(4), 100528. https://doi.org/10.1016/j.infoandorg.2024.100528

Susskind, J. (2018). *Future politics: Living together in a world transformed by tech*. Oxford University Press.

Tams, S., & And Dulipovici, A. (2024). The creativity model of age and innovation with IT: why older users are less innovative and what to do about it. *European Journal of Information Systems*, *33*(3), 287-314. https://doi.org/10.1080/0960085X.2022.2137064

Van Der Heijden, H. (2004). User Acceptance of Hedonic Information Systems. *MIS Quarterly*, *28*(4), 695-704. https://doi.org/10.2307/25148660

Wedel, M., Bigné, E., & Zhang, J. (2020). Virtual and augmented reality: Advancing research in consumer marketing. *International Journal of Research in Marketing*, *37*(3), 443-465. https://doi.org/10.1016/j.ijresmar.2020.04.004

Wu, Y., & Kelly, R. M. (2020). Online dating meets artificial intelligence: How the perception of algorithmically generated profile text impacts attractiveness and trust. Australian Conference on Human-Computer Interaction,

Yoo, Y. (2010). Computing in Everyday Life: A Call for Research on Experiential Computing. *MIS Quarterly*, *34*(2), 213-231. https://doi.org/10.2307/20721425